# Compact Modeling of Oxide-Semiconductor, 2D Material, Carbon Nanotube, and Cryogenic Transistors with Experiment Verification

Chien-Ting Tung, *Member, IEEE*

***Abstract*— This paper presents a unified compact model for emerging transistor technologies, including oxide-semiconductor field-effect transistors (OSFETs), 2D material FETs (2DFETs), carbon nanotube FETs (CNFETs), and cryogenic MOSFETs. A unified charge-density formulation is developed to account for quantum confinement, trap charges, and band-tail states in channel charge calculations. A physics-based transport model is introduced to seamlessly capture carrier transport from the long-channel diffusive regime to the short-channel ballistic limit. Scaling models are incorporated to accurately describe 2D electrostatic effects. Cryogenic operation is modeled through the inclusion of band-tail states and temperature-dependent mobility and threshold voltage. The proposed model is validated against experimental data from the fabricated OSFETs with multiple channel lengths and published measurements of 2DFETs, CNFETs, and cryogenic MOSFETs. Excellent agreement is demonstrated across diverse device architectures, operating conditions, and material systems.**

## I. Introduction

Emerging transistors have been a major driver of semiconductor technology. Although front-end-of-line (FEOL) logic devices are expected to remain dominated by silicon in the foreseeable future, emerging transistors based on new materials are poised to play an important role in three-dimensional integrated circuits (3DICs) and back-end-of-line (BEOL) applications because of their process compatibility. Amorphous oxide-semiconductor (AOS) materials, such as indium gallium zinc oxide (IGZO) and $In_2O_3$, have been widely adopted in display technologies and gain-cell memories owing to their optical transparency and low leakage current [1], [2]. More recently, $In_2O_3$ FETs have demonstrated low series resistance, high mobility, and excellent short-channel control, establishing oxide-semiconductor field-effect transistors (OSFETs) as a promising device technology for BEOL-compatible monolithic 3D integration [3].

In parallel, 2D materials and carbon nanotubes have been extensively investigated as potential alternatives to silicon technology because their atomically thin bodies can provide enhanced electrostatic control for continued channel-length scaling [4], [5], [6], [7]. Although these technologies have not yet delivered the promised ideal electrostatic behavior and are unlikely to replace silicon in the near term, they remain promising candidates for BEOL applications. They can be synthesized at relatively low temperatures compatible with BEOL thermal budgets. Although compact models for these devices have been reported [8], [9], [10], [11], [12], [13], no industry-standard compact model has yet been established for OSFETs, 2D material FETs (2DFETs) or carbon nanotube FETs (CNFETs), hindering technology pathfinding and design-technology co-optimization (DTCO) for these emerging technologies.

In this work, a unified compact model for OSFETs, 2DFETs, and CNFETs is developed. Building on the diffusive-to-ballistic framework in [14], the charge model is generalized to account for 1D/2D DOS, quantum capacitance, trap charges, and band-tail states, enabling its application to diverse material systems and cryogenic operation. Key physical parameters are constrained by experimentally reported or theoretically expected values. The remainder of this paper is organized as follows. In Section II, the unified compact model is derived. In Section III, fabricated $In_2O_3$ FETs are presented, and the model is validated for OSFETs. In Section IV, the model is verified using published data for 2DFETs, CNFETs, and cryogenic MOSFETs. In Section V, circuit-level performance predictions of emerging transistors are demonstrated using the proposed compact model.

## II. Unified Compact Model

### *A. Unified Charge Density Model*

In this work, a unified charge density model is developed to include band-tail states, trap charges, and more general quantum-confinement effects. For the drift-diffusion (DD) charge density under the gradual-channel approximation (GCA), the simplified Poisson equation can be written as (1), where $V_{gi}$ is the intrinsic gate voltage, $V_c$ is the channel quasi-Fermi potential, $V_T$ is the threshold voltage, $q_i$ is the mobile charge density, $q_Q$ is the quantum capacitance charge

The name of the corresponding author appears after the financial information, e.g. *(Corresponding author: Chien-Ting Tung)*.

*Chien-Ting Tung* is with Synopsys, Sunnyvale, CA 94085 USA. This work was conducted independently and does not represent the views or intellectual property of Synopsys.

contribution, $q_{tr}$ is the trap charge density, $\phi$ is the effective surface potential, $C_{ins}$ is the gate insulator capacitance, $C_Q$ is the quantum capacitance, $C_{tr}$ is the equivalent trap capacitance, $C_{eff}$ is the effective capacitance donated by $({C_{ins}}^{-1}+{C_Q}^{-1})^{-1}$, $\phi_{eff}$ is the effective thermal voltage considering the band-tail states, $n$ is the ideality factor, $n_q$ is the nonlinearity factor of the DOS, and $\phi_{tr}$ is the trap voltage. $q_Q$ represents the charge contribution associated with the finite quantum capacitance of the channel. Above threshold, as the carrier density increases, a portion of the applied gate voltage drops across the quantum capacitance rather than the gate insulator. The nonlinear dependence of $q_Q$ on $\phi$ allows this voltage partitioning to account for the dimensionality-dependent DOS. When $n_q$=1, it reduces to a linear quantum capacitance. $q_{tr}$ is modeled by a standard exponential charge form with $C_{tr}$ and $\phi_{tr}$ instead. $\phi$ is solved iteratively to obtain $q_i$, $q_Q$, and $q_{tr}$.

$$V_{gi}-V_c-V_T-\phi=\frac{q_i}{C_{ins}}+\frac{q_Q}{C_{ins}}+\frac{q_{tr}}{C_{eff}} \tag{1a}$$

$$q_i=C_{ins}n\phi_{eff}e^{\frac{\phi}{n\phi_{eff}}} \tag{1b}$$

$$q_Q=\frac{C_{ins}^2}{C_Q}n\phi_{eff}e^{\frac{n_q\phi}{n\phi_{eff}}} \tag{1c}$$

$$q_{tr}=C_{tr}\phi_{tr}e^{\frac{\phi}{\phi_{tr}}} \tag{1d}$$

At cryogenic temperatures, carrier transport is strongly influenced by band-tail states. A rigorous evaluation of the corresponding band-tail charge generally requires the Fermi–Dirac distribution, which does not yield a closed-form solution. To obtain a compact and computationally efficient formulation, exponential band-tail states with a band-tail temperature/voltage are commonly used to describe the mobile charge and have been shown to provide accurate results in compact modeling [15], [16]. Rather than empirically clamping the thermal voltage using this band-tail temperature [15], a physically motivated function for $\phi_{eff}$ is introduced. The mobile charge can be expressed as the sum of the DOS charge and the band-tail charge, as given in (2a), where $\phi_t$ is the thermal voltage and $\phi_{tail}$ is the band-tail voltage. The band-tail states mostly affect the subthreshold region so we will focus on the regio where -∞ $<\phi<0$. Integrating (2a) from -∞ to 0, we obtain $\phi_t+\phi_{tail}$ as (2b), from which we can get $\phi_{eff}$ be equal to $\phi_t+\phi_{tail}$. For compact modeling purposes, a smoothing factor $n_{tail}$ is introduced in (2c) to describe the transition between $\phi_t$ and $\phi_{tail}$.

$$q_i\propto e^{\frac{\phi}{\phi_t}}+e^{\frac{\phi}{\phi_{tail}}} \tag{2a}$$

$$\int_{-\infty}^{0}e^{\frac{\phi}{\phi_t}}+e^{\frac{\phi}{\phi_{tail}}}d\phi=\phi_t+\phi_{tail}=\int_{-\infty}^{0}e^{\frac{\phi}{\phi_{eff}}}d\phi \tag{2b}$$

$$\phi_{eff}=(\phi_t^{n_{tail}}+\phi_{tail}^{n_{tail}})^{1/n_{tail}} \tag{2c}$$

To account for short-channel effects, both $V_T$ and $n$ are modeled as length dependent quantities. Previous studies have shown that $V_T$ roll-off, drain-induced-barrier-lowering (DIBL), and subthreshold swing (SS) degradation follow exponential dependences on the channel length ($L$) as (3) and (4) [17], where $V_{T0}$ is the long-channel threshold voltage, dvt0 is the $V_T$ roll-off factor, $\eta_0$ is the long-channel DIBL factor, $d\eta$ is the DIBL roll-off factor, $l_{sc1}$ is the scaling length for $V_T$, $n_0$ is the long-channel ideality factor, dn0 is the SS degradation factor, $n_d$ is the punch-through factor, dnd is the punch-through degradation factor, and $l_{sc2}$ is the scaling length for $n$. $V_{dsi}$ is the intrinsic drain-source voltage. Although not explicitly shown here, $|V_{dsi}|$ is implemented as a smoothed absolute-value function of $V_{dsi}$, as presented in [14]. All $|x|$ in this paper are represented using smoothed absolute-value functions to preserve continuity.

$$V_T=V_{T0}-\frac{\mathrm{dvt0}}{\exp\left(\frac{L}{l_{sc1}}\right)-1}-\left(\eta_0+\frac{d\eta}{\exp\left(\frac{L}{l_{sc1}}\right)-1}\right)|V_{dsi}| \tag{3}$$

$$n=n_0+\frac{\mathrm{dn0}}{\exp\left(\frac{L}{l_{sc2}}\right)-1}+\left(n_d+\frac{\mathrm{dnd}}{\exp\left(\frac{L}{l_{sc2}}\right)-1}\right)|V_{dsi}| \tag{4}$$

For the ballistic transport (BT), the transmission factor $\mathcal{T}=\lambda/(\lambda+L)$ is introduced to calculate the charge at top-of-the-barrier ($q_{\mathrm{top}}$, $q_{\mathrm{Qtop}}$, and $q_{\mathrm{trtop}}$) as (5). $\lambda$ is the mean-free-path (MFP) of the channel. $V_{top}$ is the channel Fermi potential at the virtual source (VS).

$$V_{gi}-V_{top}-V_T-\phi=\frac{q_{top}}{C_{ins}}+\frac{q_{Qtop}}{C_{ins}}+\frac{q_{trtop}}{C_{eff}} \tag{5a}$$

$$q_{top}=\frac{2-\mathcal{T}+\mathcal{T}e^{-\frac{|V_{dsi}|}{\phi_{eff}}}}{2}C_{ins}n\phi_{eff}e^{\frac{\phi}{n\phi_{eff}}} \tag{5b}$$

$$q_{Qtop}=\frac{2-\mathcal{T}+\mathcal{T}e^{-\frac{|V_{dsi}|}{\phi_{eff}}}}{2}\frac{C_{ins}^2}{C_Q}n\phi_{eff}e^{\frac{n_q\phi}{n\phi_{eff}}} \tag{5c}$$

$$q_{trtop}=\frac{2-\mathcal{T}+\mathcal{T}e^{-\frac{|V_{dsi}|}{\phi_{eff}}}}{2}C_{tr}\phi_{tr}e^{\frac{\phi}{\phi_{tr}}} \tag{5d}$$

## *B. Current Model*

The channel current ($I_{ds}$) model incorporates both DD and BT, following the previous work [14]. $I_{ds}$ is described as $q_{\mathrm{top}}$ multiplied by the effective carrier velocity ($v_{eff}$), which is limited by the thermal velocity ($v_{th}$) through a saturation function as (6) where $W$ is the effective width of the channel.

$$I_{ds}=Wq_{top}\frac{v_{eff}}{\left(1+\left(\frac{|v_{eff}|}{v_{th}}\right)^4\right)^{1/4}} \tag{6a}$$

$$v_{eff}=\frac{v_{bt}v_{dd}}{v_{bt}+v_{dd}} \tag{6b}$$

$v_{bt}$ is the ballistic velocity defined as $\frac{v_{th}V_{dsi}}{2\phi_{eff}}$. The DD velocity ($v_{dd}$) is obtained by integrating (7) using (1), yielding (8), where $\mu_{eff}$ is the effective mobility, $\mu_0$ is the low-field mobility, and UA and EU are the fitting parameters for mobility degradation. The length dependent saturation velocity $v_{sat}$ is adopted from [14], [18], where $v_{sat0}$ is the long-channel saturation velocity and $\lambda_{sat}$ is the scaling factor. $\beta_1$ is a fitting parameter for velocity saturation. $q_s$ and $q_d$ are the DD mobile charge densities at the source and drain ends obtained from (1), whereas $q_{trs}$ and $q_{trd}$ are the DD trap charge densities. $\phi_{sd}$ is the effective channel potential difference, and $V_{dsat1}$ is equal to $v_{sat}L/\mu_{eff}$. The first two numerator terms in (8a) correspond to the conventional drift and diffusion contributions of the standard MOSFET formulation, whereas the third and fourth terms are additional corrections arising from quantum capacitance and trap charge, respectively.

$$\frac{I_{ds}L}{W} = \int \mu_{eff} q_i \frac{dV_c}{dq_i} dq_i \tag{7}$$

$$v_{dd} = \frac{\mu_{eff}}{q_{\mathrm{top}}L} \times \frac{\frac{q_s^2-q_d^2}{2C_{ins}} + n\phi_{eff}(q_s-q_d) + \frac{n_q\left(q_s^{n_q+1}-q_d^{n_q+1}\right)}{(n_q+1)C_Q(C_{ins}n\phi_{eff})^{n_q-1}} + \frac{n\phi_{eff}(q_s q_{trs}-q_d q_{trd})}{(n\phi_{eff}+\phi_{tr})C_{eff}}}{\left(1+\left(\frac{|\phi_{sd}|}{V_{dsat1}}\right)^{\beta_1}\right)^{1/\beta_1}} \tag{8a}$$

$$\phi_{sd} = \frac{q_s-q_d}{C_{ins}} + \frac{\left(q_s^{n_q}-q_d^{n_q}\right)}{C_Q(C_{ins}n\phi_{eff})^{n_q-1}} + \frac{q_{trs}-q_{trd}}{C_{eff}} \tag{8b}$$

$$\mu_{eff} = \frac{\mu_0}{1+\mathrm{UA}\left(q_{top}/(0.0259C_{eff})\right)^{\mathrm{EU}}} \tag{8c}$$

$$v_{sat} = v_{sat0}\left(1+\frac{\lambda_{sat}}{L}\right) \tag{8d}$$

To solve (1), $V_c$ is required. Owing to velocity saturation, the charge density must remain above a minimum value. For a given $q_{top}$, $q_{min}$ is determined from current continuity and the DD current expression in (8a), leading to (9a) and (9b). The saturation voltage ($V_{dsat2}$) is then obtained from $q_{topx}$ and $q_{min}$ as (9c), where $q_{trtopx}$ and $q_{trmin}$ are the trap charge densities corresponding to $q_{topx}$ and $q_{min}$. Finally, the $V_c$ at the source and drain ends are computed using (9d) and (9e) with $\beta_2$ as a fitting parameter. The superscripts $d$ and $s$ donate the source- and drain-end positions in the channel, respectively, and $V_{si}$ and $V_{di}$ are the intrinsic source and drain voltage.

$$q_{topx} = 2q_{top}/\left(2-\mathcal{T}+\mathcal{T}exp\left(-|V_{dsi}|/\phi_{eff}\right)\right) \tag{9a}$$

$$q_{min} = q_{topx} - C_{eff}V_{dsat1}\left(\sqrt{\frac{2q_{topx}}{C_{eff}V_{dsat1}}}-1\right) \tag{9b}$$

$$V_{dsat2} = \frac{q_{topx}-q_{min}}{C_{ins}} + \frac{\left(q_{topx}^{n_q}-q_{min}^{n_q}\right)}{C_Q(C_{ins}n\phi_{eff})^{n_q-1}} + n\phi_{eff}\ln\left(\frac{q_{topx}}{q_{min}}\right) + \frac{q_{trtopx}-q_{trmin}}{C_{eff}} \tag{9c}$$

$$V_{dsat}^{d,s} = V_{dsat2} + e^{\mp V_{dsi}/\phi_{eff}} \tag{9d}$$

$$V_c^{d,s} = V_{si,di} \pm \frac{V_{dsi}}{\left(1+\left(|V_{dsi}|/V_{dsat}^{d,s}\right)^{\beta_2}\right)^{1/\beta_2}} \tag{9e}$$

## C. Intrinsic Charge Model

To obtain the total channel charge, current continuity must be enforced while integrating (1), which does not yield a closed-form solution. Therefore, the charge is approximated by incorporating the trap charge into the conventional MOSFET intrinsic charge model [8], [19], as given in (10). Equations (10d) and (10e) follow the conventional MOSFET formulation, except that $q_{trs}$ and $q_{trd}$ are introduced into $q_{ia}$ and $dq$.

$$q_{ia} = 0.5(q_s + q_{trs} + q_d + q_{trd}) \tag{10a}$$

$$dq = 0.5(q_s + q_{trs} - q_d - q_{trd}) \tag{10b}$$

$$K_q = 2(q_{ia} + C_{eff}n\phi_{eff}) \tag{10c}$$

$$Q_{Gi} = WL\left(q_{ia} + \frac{dq^2}{6K_q}\right) \tag{10d}$$

$$Q_{Di} = -WL\left(q_{ia} - \frac{dq}{6}\left(1-\frac{dq}{K_q}\left(1+\frac{dq}{5K_q}\right)\right)\right) \tag{10e}$$

$$Q_{Si} = -Q_{Gi} - Q_{Di} \tag{10f}$$

Equation (10) models the charge under DD conditions. For BT and quasi-BT operation, several modifications are required. The charge densities at the source and drain ends are expressed as linear combinations of the DD charge ($q_{DD}$) and BT charge ($q_{BT}$) as (11a). A multiplication factor $K_{bt}$, defined in (11b), is applied to (10d) and (10e) for the nonequilibrium charge distribution in the channel, where $\theta$ is a fitting parameter. The detailed derivation and implementation are provided in our previous work [14].

$$(1-\mathcal{T})q_{DD} + \mathcal{T}q_{BT} \tag{11a}$$

$$K_{bt} = \sqrt{1+\theta|V_{dsi}|} \tag{11b}$$

## D. Temperature Dependence Model

To maintain model conciseness, a minimum set of temperature-dependent parameters is selected to extend the model from room temperature to cryogenic operation. The threshold voltage is temperature dependent because of the temperature dependence of the bandgap. The thermal velocity, mobility, and saturation velocity are also temperature dependent owing to their dependence on thermal energy and scattering probability. In addition, $\beta_2$ is made temperature dependent to capture the observed shift in the drain-current saturation knee with temperature. The temperature-dependent key model parameters are therefore calculated using (12), where $V_{T0,nom}$, $v_{th,nom}$, $v_{sat0,nom}$, and $\beta_{2,nom}$ are the parameters for the nominal temperature while dvtt, vtht, ute, vsatt, and beta2t are the temperature dependence fitting parameters. $T$ is the device temperature.

$$V_{T0} = V_{T0,nom} + \mathrm{dvtt}\left(1-\frac{T}{300}\right) \tag{12a}$$

$$v_{th} = v_{th,nom}\left(\frac{T}{300}\right)^{\mathrm{vtht}} \tag{12b}$$

$$\mu_0 = \frac{v_{th}\lambda}{2\phi_t}\left(\frac{T}{300}\right)^{\mathrm{ute}} \tag{12c}$$

$$v_{sat0} = v_{sat0,nom} + \mathrm{vsatt}\left(1-\frac{T}{300}\right) \tag{12d}$$

$$\beta_2 = \beta_{2,nom} + \mathrm{beta2t}\left(\frac{T}{300}-1\right) \tag{12e}$$

## III. $In_2O_3$ FET Experiment and Model Validation

To validate the proposed model for OSFETs, measured data from fabricated enhancement-mode $In_2O_3$ FETs were provided by Dr. Shao of MIT [3]. The key fabrication steps are summarized as follows. First, a tungsten (W) gate was sputter-deposited on a silicon-on-insulator (SOI) substrate. A 4.8-nm-thick $HfO_2$ gate dielectric was then formed by plasma-enhanced atomic layer deposition (PEALD) at 250 °C, followed by the deposition of a 2.3-nm-thick $In_2O_3$ channel layer by PEALD at 150 °C. A Ni/Au bilayer was used for the source/drain contacts. The fabricated devices had channel lengths ranging from 40 to 970 nm. The device schematic, process flow, and scanning electron microscopy (SEM) image are shown in Fig. 1.

First, the charge- and capacitance-related parameters are extracted from the IdVg curve at $V_{ds}$ = 0.05 V and the CV data of the 970-nm transistor, as shown in Fig. 2. A key difference between AOS and crystalline silicon is the trap density. Compared with silicon, AOS exhibits a substantially higher trap density, which cannot be neglected. Fig. 2 compares the model extracted with and without trap charge. When trap charge is included, the weak-inversion region exhibits a smoother transition and agrees well with the measured data, in contrast to the model without trap charge. The fitting shows a trap voltage of 0.0604 V which is equivalent to a 700 K trap temperature.

Next, the remaining model parameters are extracted using the IdVg and IdVd characteristics of devices with channel lengths of 970, 470, and 40 nm. To determine $v_{th}$, a commonly reported apparent effective mass of $0.3m_0$ is selected [20], resulting in a velocity of $10^7$ cm/s, consistent with the value measured in [3]. With this $v_{th}$, the extracted model shows excellent agreement with the measured data, as shown in Figs. 3–5. A single set of fitting parameters is used, except for $C_{tr}$, demonstrating the scalability of this physical model. The extracted parameters are summarized in Table I. No temperature dependence fitting parameter is used. $R_s$ and $R_d$ are the source and drain series resistances. The MFP of these devices is 1.7 nm, as expected for thin-film amorphous materials; therefore, transport remains DD dominated even at a channel length of 40 nm. The ultrathin-body structure of these $In_2O_3$ FETs provides excellent gate control, with zero DIBL roll-off and nearly linear $V_T$ roll-off extracted from the measurements.

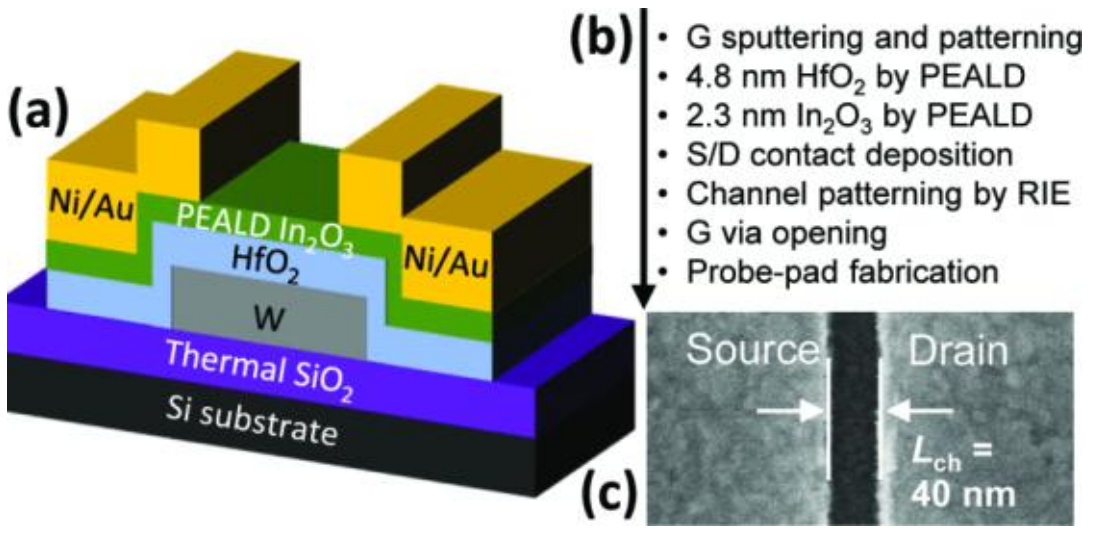

Fig. 1. (a) The device schematic, (b) the process flow, and (c) scanning electron microscopy (SEM) of $In_2O_3$ FETs [3] ©2026 IEEE. Reprinted with permission of the author.

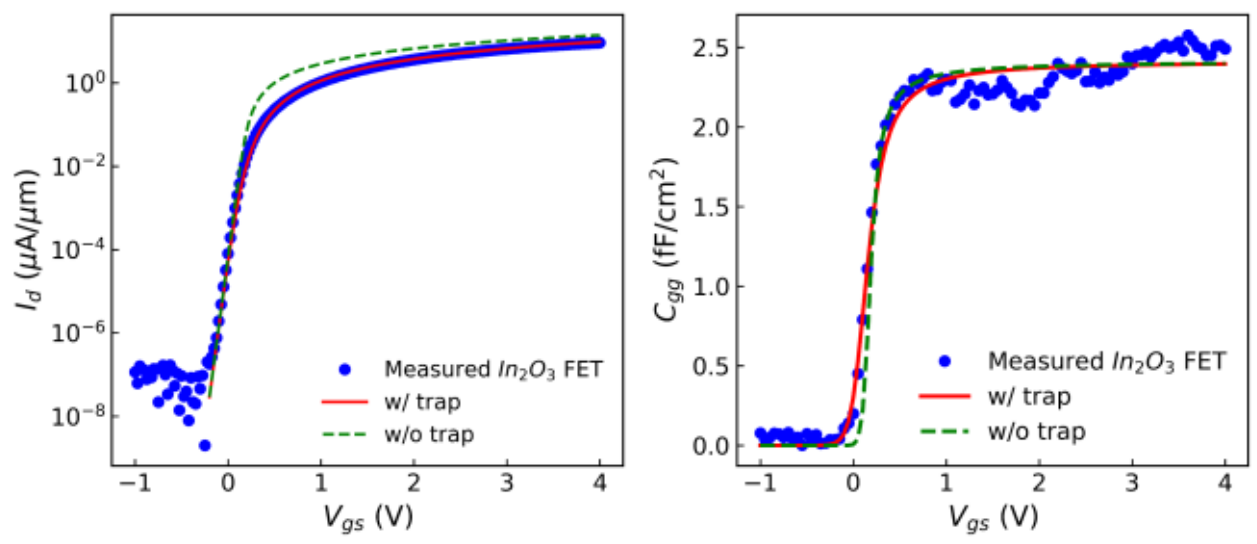

Fig. 2. Measured [3] and modeled IdVg ($V_{ds}$=0.05 V) and CV of the 970-nm $In_2O_3$ FET.

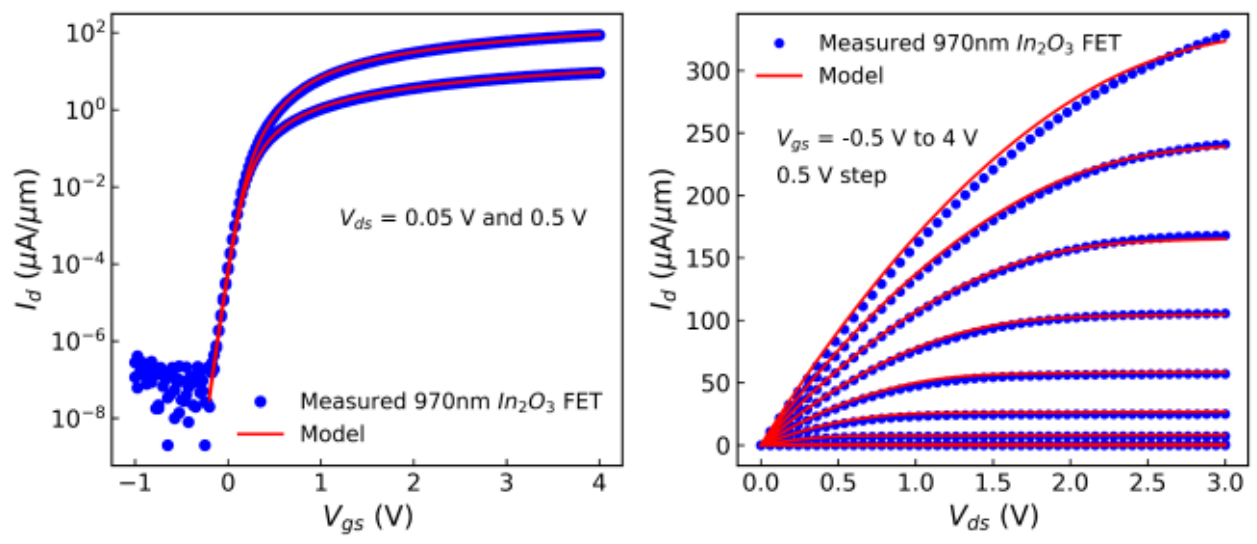

Fig. 3. Measured [3] and modeled IdVg and IdVd of the 970-nm $In_2O_3$ FET.

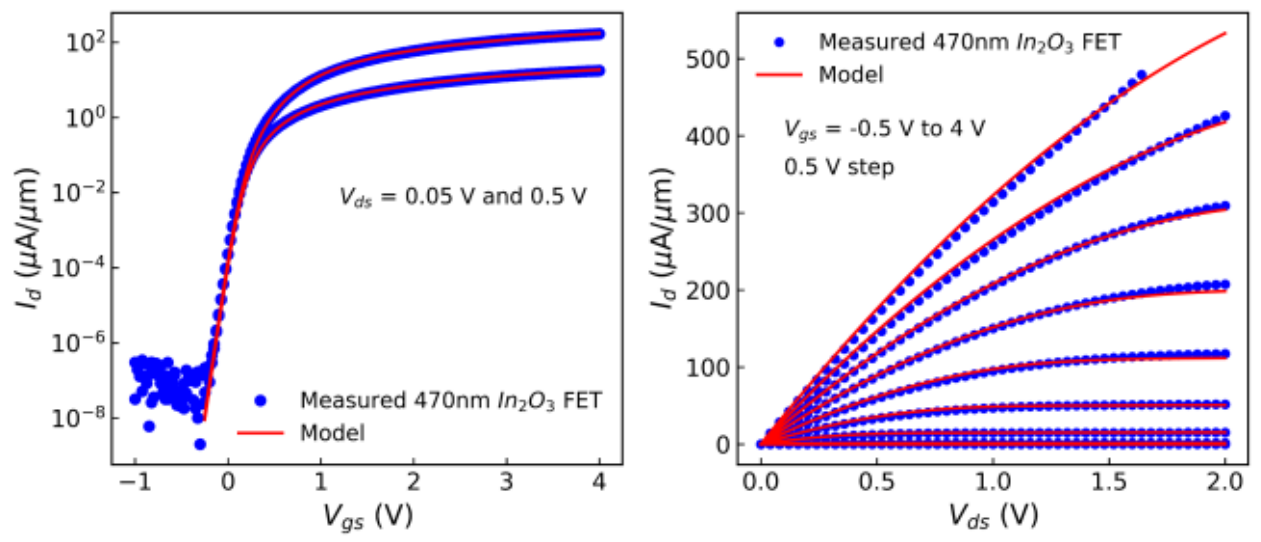

Fig. 4. Measured [3] and modeled IdVg and IdVd of the 470-nm $In_2O_3$ FET.

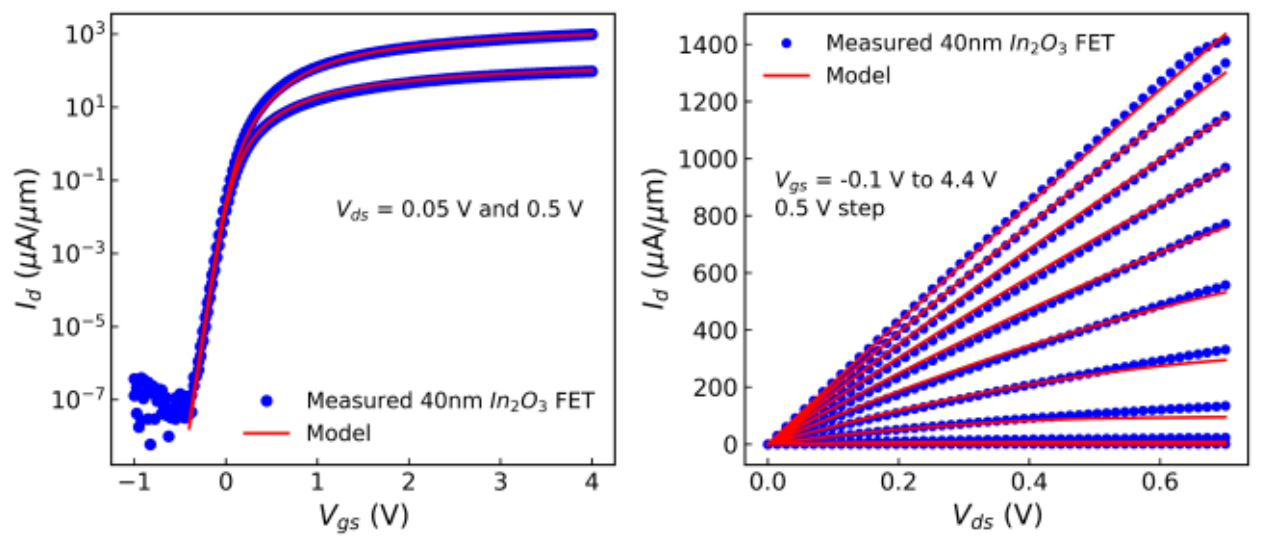

Fig. 5. Measured [3] and modeled IdVg and IdVd of the 40-nm $In_2O_3$ FET.

TABLE I
FITTING PARAMETERS FOR $In_2O_3$ FETS

| | | | |
|---|---|---|---|
| $C_{ins}$ | $C_Q$ | $n_q$ | $\phi_{tr}$ |
| 0.037 F/m² | 0.07 F/m² | 1 | 0.0604 V |
| $v_{th}$ | $\lambda$ | $v_{sat0}$ | $\lambda_{sat}$ |
| $10^7$ cm/s | 1.7 nm | $10^7$ cm/s | 1.7 nm |
| $\beta_1$ | $\beta_2$ | $V_{T0}$ | dvt0 |
| 4 | 8 | 0.18 V | 0.1 V |
| $l_{sc1}$ | $n_0$ | dn0 | $l_{sc2}$ |
| 50 nm | 1 | 0.05 | 50 nm |
| $l_{sc2}$ | $\eta_0$ | $d\eta$ | $n_d$ |
| 50 nm | 0.01 | 0 | 0 |
| dnd | UA | EU | $R_s$ & $R_d$ |
| 0 | 0 | 0 | 120 Ω·μm |
| $C_{tr}$ (970 nm) | $C_{tr}$ (470 nm) | $C_{tr}$ (40 nm) | |
| 0.08 F/m² | 0.085 F/m² | 0.12 F/m² | |

## IV. 2DFET, CNFET, AND CRYOGENIC MOSFET MODEL VALIDATION

### A. 2DFET

2DFETs, such as $MoS_2$ FETs, have been investigated for many years. The atomically thin body of 2DFETs makes quantum capacitance more important than in conventional MOSFETs, which could reduce the inversion capacitance. In this model, the 2D DOS is represented by $C_Q$ with a theoretical $n_q$ value of 1, corresponding to a linear capacitance [21]. Contact resistance has been a major obstacle for $MoS_2$ FETs. Recently, yttrium-induced phase-transition $MoS_2$ FETs have demonstrated nearly ideal ohmic contact [4]. In this process, trilayer $MoS_2$ is used as the channel, and 2.6-nm $HfO_2$ is used as the gate insulator. To model this device, the effective mass of $MoS_2$ is taken as $0.38m_0$, as reported in [22]. Accordingly, $v_{th}$ is $8\times10^6$ cm/s, and the 2D quantum capacitance is 0.255 F/m. Based on these parameters, the remaining parameters are extracted to fit the measured characteristics of the 10-nm $MoS_2$ FET, as shown in Fig. 6. The extraction shows a good contact resistance of 85 Ω·µm. The MFP of this device is 2.3 nm, which indicates that $MoS_2$ FETs remain diffusive even below 10 nm.

## B. CNFET

Carbon nanotubes are widely considered an ultimate channel material because their one-dimensional nanowire structure provides excellent electrostatic control. Unlike 2D materials, whose DOS yields an approximately constant quantum capacitance, the 1D DOS of carbon nanotubes results in a nonlinear quantum capacitance. This behavior is captured in the proposed model through $n_q$. Fig. 7 shows the reported [12] and modeled intrinsic gate capacitance of a CNFET. When quantum capacitance is neglected, the model predicts a conventional CV characteristic, as shown by the green curve. Including quantum charge with $n_q = 2.6$ introduces the nonlinear quantum-capacitance effect, causing a larger fraction of the channel potential to be consumed and thereby reducing the total gate capacitance. The model accurately reproduces the reported CNFET capacitance behavior.

Using the calibrated $n_q$ value for carbon nanotubes, the proposed model was fitted to the measured CNFET data with channel lengths of 90 and 10 nm [6]. A $v_{th}$ of $4.1 \times 10^5$ m/s was extracted from the measurements and used in the fitting. Because the measured data do not exhibit the expected $V_T$ roll-off, indicating process variability, separate parameter sets were extracted for the two devices. The fitting results and extracted parameters are presented in Figs. 8 and 9. The extracted $WC_{ins}$ is consistent with the value reported in [6]. The extracted parameters are nearly identical for the 10- and 90-nm CNFETs except for the short-channel-effect parameters, once again, showing the good scalability of the proposed model. The extracted MFP also agrees well with reported values for short-channel CNFETs in [6]. The fitting indicates source/drain resistances of 5.7 and 8 kΩ also matches the values from [6]. Comparing to the 10-nm $MoS_2$ FET, 10-nm CNFET experiences a much higher DIBL which limits its potential applications in high-performance computing and requires more studies on process engineering.

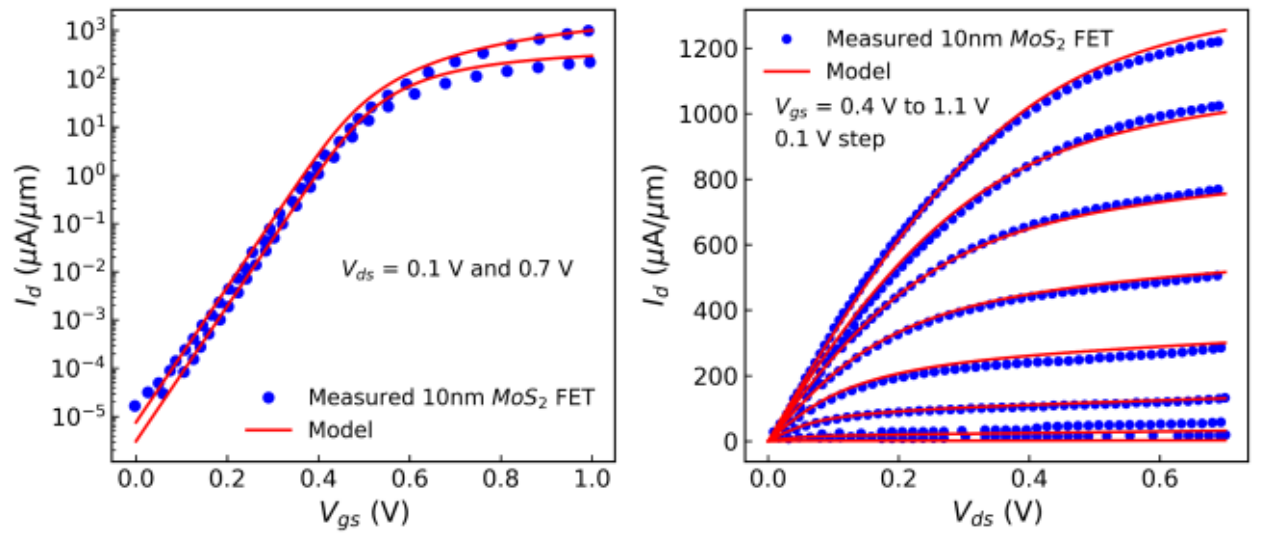

Fig. 6. Measured [4] and modeled IdVg and IdVd of the 10-nm $MoS_2$ FET. The extracted parameters are $C_{ins}$=0.068 F/m$^2$, $C_Q$=0.255 F/m$^2$, $n_q$=1, $V_{T0}$=0.49 V, $v_{th}$=8×10$^6$ cm/s, $\lambda$=2.3 nm, $v_{sat0}$=8×10$^6$ cm/s, $\lambda_{sat}$=2.3 nm, $\beta_1$=4, $\beta_2$=1.5, $n_0$=1.2, $n_d$=0, $\eta_0$=0.04, UA=0, EU=0, and $R_s$ & $R_d$=85 Ω·µm.

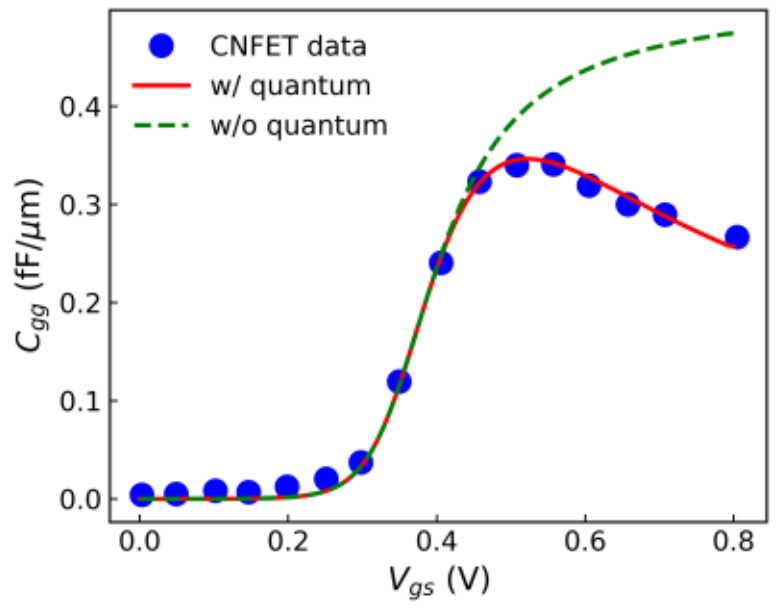

Fig. 7. Capacitance data [12] of the CNFET versus the model's simulation. $WC_{ins}$=0.523 fF/µm, $WC_Q$=44.2 fF/µm, and $n_q$=2.6 are used in the fitting.

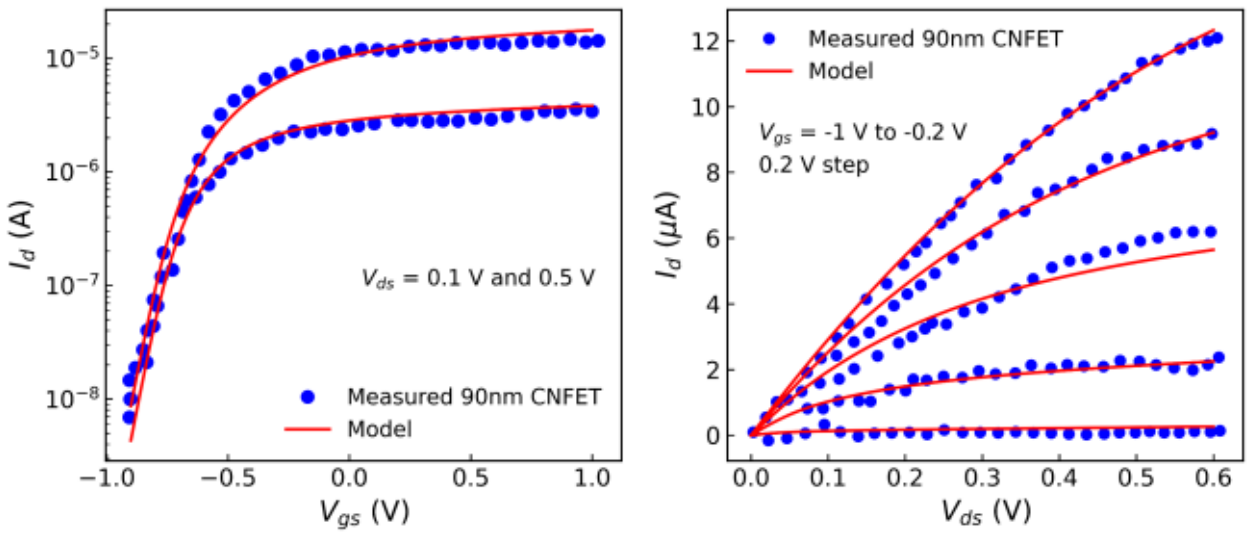

Fig. 8. Measured [6] and modeled IdVg and IdVd of the 90-nm CNFET. The extracted parameters are $WC_{ins}$=0.177 fF/µm, $WC_Q$=13.6 fF/µm, $n_q$=2.6, $V_{T0}$=-0.75 V (IdVg), $V_{T0}$=-1 V (IdVd), $v_{th}$=4.1×10$^7$ cm/s, $\lambda$=10 nm, $v_{sat0}$=4.1×10$^7$ m/s, $\lambda_{sat}$=10 nm, $\beta_1$=4, $\beta_2$=1, $n_0$=1.5, $n_d$=0, $\eta_0$=0.05, and $R_s$ & $R_d$=8 kΩ. Due to the inconsistency of the data, IdVg and IdVd fitting use different $V_{T0}$.

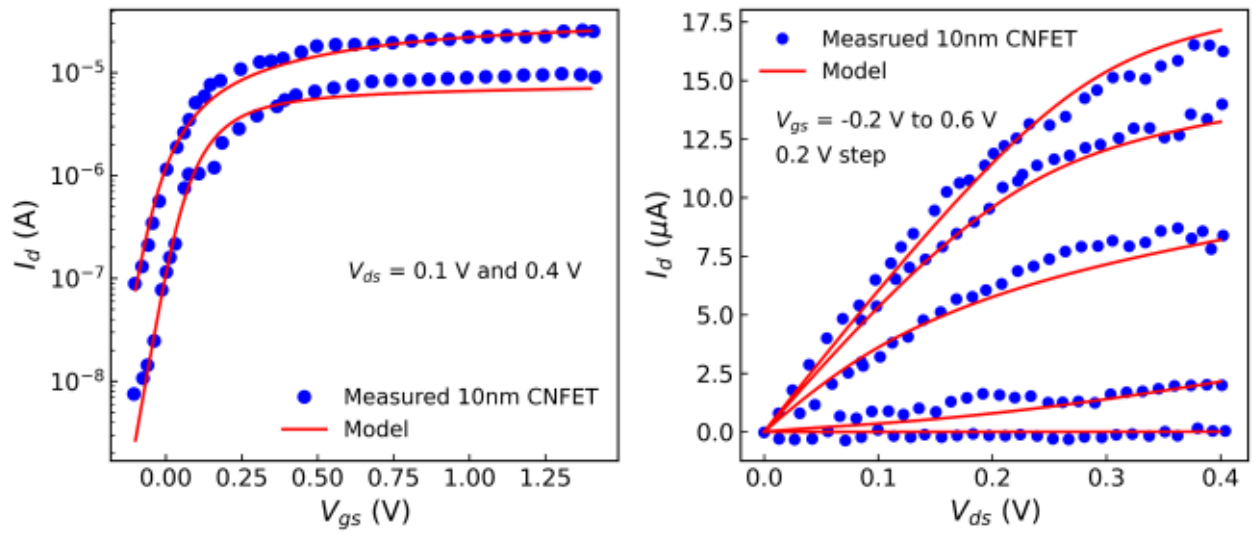

Fig. 9. Measured [6] and modeled IdVg and IdVd of the 10-nm CNFET. The extracted parameters are $WC_{ins}$=0.177 fF/µm, $WC_Q$=13.6 fF/µm, $n_q$=2.6, $V_{T0}$=0.06 V (IdVg), $V_{T0}$=0.02 V (IdVd), $v_{th}$=4.1×10$^7$ cm/s, $\lambda$=10 nm, $v_{sat0}$=4.1×10$^7$ m/s, $\lambda_{sat}$=10 nm, $\beta_1$=4, $\beta_2$=1, $n_0$=1, $n_d$=0, $\eta_0$=0.28, and $R_s$ & $R_d$=5.7 kΩ. Due to the inconsistency of the data, IdVg and IdVd fitting use different $V_{T0}$.

## C. Cryogenic MOSFET

The previous validation results did not assess the band-tail-state formulation or the temperature-dependent model. Therefore, the model was further validated against measured FinFET data from 300 K to 8 K [23], as shown in Fig. 10. The device is a long-channel FinFET with a gate length of 130 nm, an oxide thickness of 2.5 nm, and a total fin width of 150 nm. The measured data exhibit SS saturation below 50 K, corresponding to a $\phi_{tail}$ of 4.3 mV. No quantum-capacitance effect is observed because of the relatively large fin thickness (30 nm). The proposed model accurately reproduces the measured characteristics from room temperature to cryogenic temperatures with the parameters summarized in TABLE II.

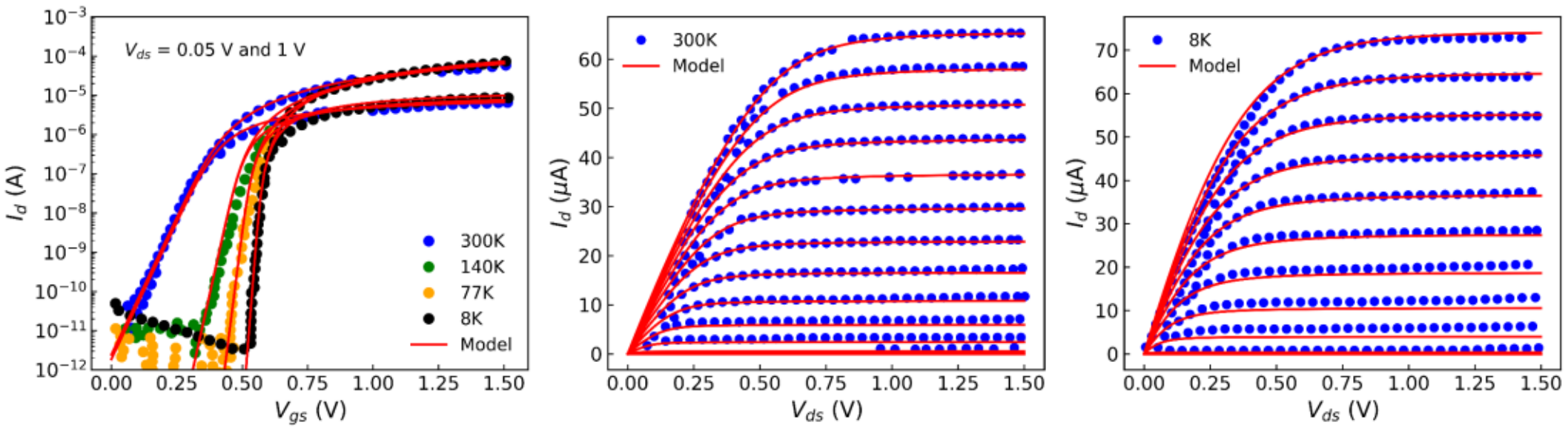

Fig. 10. Measured [23] and modeled IV curves of a FinFET. IdVg is from 300K to 8K. IdVd is at 300K and 8K.

TABLE II
FITTING PARAMETERS FOR CRYOGENIC FINFET

| $C_{ins}$ | $\phi_{tail}$ | $V_{T0}$ | $v_{th}$ |
|---|---|---|---|
| 0.0138 F/m$^2$ | 4.3 mV | 0.37 V | 1.14×10$^7$ cm/s |
| $\lambda$ | $v_{sat0}$ | $\lambda_{sat}$ | $\beta_1$ |
| 11 nm | 5.5×10$^6$ cm/s | 11 nm | 4 |
| $\beta_2$ | $n_0$ | UA | EU |
| 8 | 1.2 | 3.6×10$^{-2}$ | 1 |
| $\eta_0$ | $n_d$ | dvtt | ute |
| 0 | 0 | 0.2 V | 0.86 |
| dvsatt | beta2t | vtht | $R_s$ & $R_d$ |
| 8×10$^5$ cm/s | 2.4 | 0 | 500 Ω |

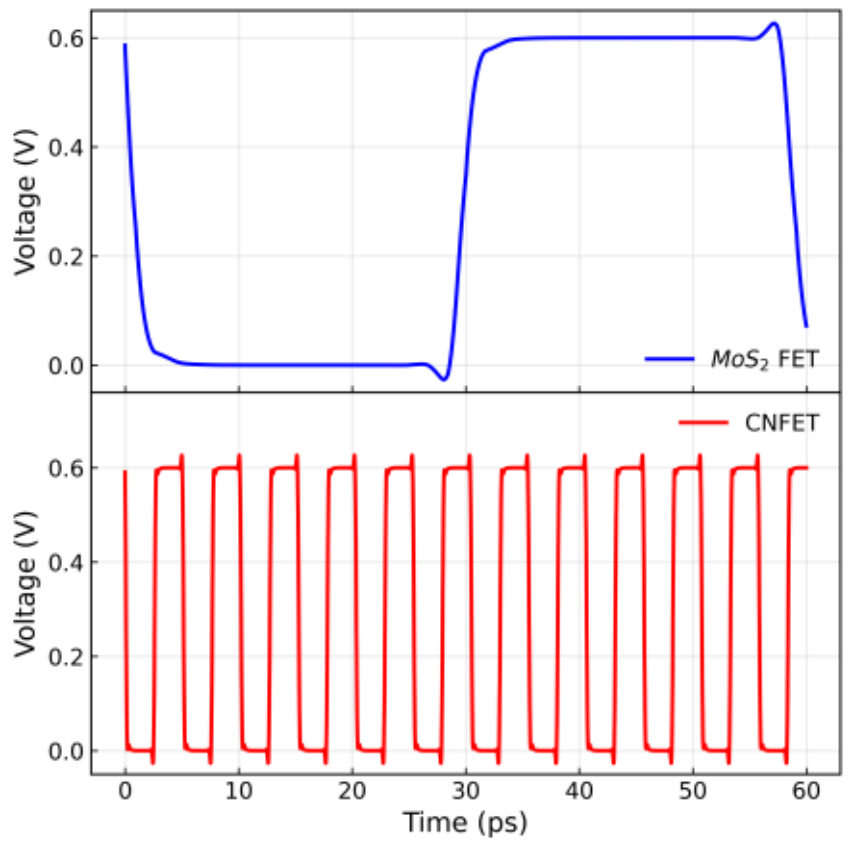

Fig. 11. 17-stage ring oscillator simulations for the calibrated $MoS_2$ FET and CNFET.

## V. PERFORMANCE PREDICTION OF EMERGING TRANSISTORS

Using the calibrated models, the transistor performance is predicted through circuit-level simulations. The models are implemented in Verilog-A and simulated in Ngspice [24]. To estimate the upper bound of device performance, this section considers only the intrinsic device characteristics and excludes parasitic components. The analysis is intended to evaluate the potential of $MoS_2$ FETs and CNFETs for future advanced technology nodes beyond BEOL. A fanout-of-3, 17-stage ring oscillator is used to evaluate the intrinsic delay. The threshold voltages of the calibrated $MoS_2$ FET model in Fig. 6 and the CNFET model in Fig. 9 are adjusted to match the off-current target in the IRDS 2023 [25]. The supply voltage is set to 0.6 V, consistent with the 1.5-nm technology node in the IRDS 2023. The simulation results are shown in Fig. 11. The intrinsic delays are 1.7 ps for the $MoS_2$ FET and 0.15 ps for the CNFET, which represent estimated upper-bound performance values for these devices under their current process technologies. Owing to the larger effective mass and lower MFP of $MoS_2$, the $MoS_2$ FET exhibits a lower on-current and a longer delay than the corresponding IRDS target. Nevertheless, the atomically thin $MoS_2$ channel provides excellent DIBL (40 mV) at 10-nm channel length, making it suitable for ultra-scaled channel transistors and offsetting the impact of its low carrier velocity. In contrast, the CNFET achieves a substantially shorter delay because of its high thermal velocity and mobility. However, the extracted CNFET experiences a large DIBL (280 mV) implying large leakage current at high applied voltages. Improving gate control and material properties remains necessary to suppress DIBL and other short-channel effects in CNFETs.

## VI. CONCLUSION

In this work, a unified compact model has been developed for emerging transistors, including OSFETs, 2DFETs, CNFETs, and cryogenic MOSFETs. The model incorporates quantum capacitance, trap charges, band-tail states, ballistic transport, and temperature dependence. It is validated using multiple experimental datasets, with key physical parameters, including effective mass, thermal velocity, and mobility, constrained by experimentally reported or theoretically expected values. The results demonstrate that trap charges and quantum capacitance play critical roles in accurately calibrating compact models for these device technologies. The proposed model provides a predictive framework for evaluating future emerging-transistor technologies and supports technology pathfinding.

## ACKNOWLEDGMENT

The author would like to thank Yanjie Shao from MIT for providing the $In_2O_3$ FET data. Large language models were used solely for text editing and language refinement. No data, figure, or analysis was generated by AI.